# Modelling cortical network dynamics

G.K. Cooray[1,2,3], R.E. Rosch[2,4] and K.J. Friston[4]

[1] GOS-UCL Institute of Child Health, University College London, London, UK

[2] Great Ormond Street Hospital NHS Foundation Trust, London, UK

[3] Karolinska Institutet, Stockholm, Sweden

[4] The Wellcome Centre for Human Neuroimaging, Queen Square Institute of Neurology, University College London, London, UK

Email: gerald.cooray@ki.se

**Abstract**

We consider the theoretical constraints on interactions between coupled cortical columns. Each column comprises a set of neural populations, where each population is modelled as a neural mass. The existence of semi-stable states — within a cortical column — has been shown to be dependent on the type of interaction between the constituent neuronal subpopulations, i.e., the form of the implicit synaptic convolution kernels. Current-to-current coupling has been shown – in contrast to potential-to-current coupling — to create semi-stable states within a cortical column. In this analytic and numerical study, the interaction between semi-stable states is characterized by equations of motion for ensemble activity. We show that for small excitations, the dynamics follow the Kuramoto model. However, in contrast to previous work, we derive coupled equations between phase and amplitude dynamics. This affords the possibility of defining connectivity as a dynamic variable. The turbulent flow of phase dynamics — found in networks of Kuramoto oscillators — indicate turbulent changes in dynamic connectivity for coupled cortical columns. Crucially, this is something that has been recorded in epileptic seizures. We used the results we derived to estimate a seizure propagation model, which allowed for relatively straightforward inversions using variational Laplace (a.k.a., Dynamic Causal Modelling). The face validity of the ensuing seizure propagation model was established using simulated data as a prelude to future work; which will investigate dynamic connectivity from empirical data. This model also allows predictions of seizure evolution, induced by virtual lesions to synaptic connectivity: something that could be of clinical use, when applied to epilepsy surgical cases.

**Author Summary**

The human brain is covered by a thin layer, the cortex, comprising specific types of layered neuronal cells. The cortex consists of columns with internal neuronal information processing and together these columns interact, allowing for the assimilation of large amounts of information. We have previously characterized the mathematical constraints on the interactions of brain activity within a small region. In this paper, we give a theoretical description of interactions over a larger region of the brain. The resulting formalism is applied to understand how seizures in people with epilepsy propagate across the brain. Moreover, using data from seizure propagation, we can infer how different regions of the brain interact during a seizure. This could be used to tailor the surgical procedures used to treat patients with complex epilepsy.

## Introduction

The human cortex comprises a lattice of cortical columns with dense intracolumnar connections and a sparse network of intercolumnar connections [1]. The histological structure of the cortex features several layers of different types of neuronal cells, with different structures and receptors [2]. Neuronal activity within each column generates extracellular currents whose effects can be measured using electrodes and sensors in the near vicinity, and also more coarsely — at greater distances – outside the human body as recorded with scalp electro- and magneto-encephalography (EEG and MEG) [3]. These recordings reflect broadband frequency activity: the frequency content is shaped by the different types of postsynaptic receptors present in the source of cortical activity. Cortical activity also reports stochastic features of the underlying dynamics, including spikes, sharp transients, and paroxysmal rhythms, seen both in the healthy and dysfunctional brain [3].

Key features of the dynamics of cortical column activity have been described by a multitude of generative models, ranging from one-dimensional integrate and fire neurons, multidimensional neuronal mass models and infinite dimensional partial differential equations. Neuronal mass models allow for sufficient simplification of the cortical columns to allow for analytically tractable dynamics of cortical networks [4-6]. The simplest neural mass models are the convolution-based models, where synaptic kernels are used to convolve presynaptic input to produce postsynaptic dynamics. The canonical microcircuit model typifies a convolution-based neural mass model of a cortical column and comprises 4 interacting excitatory/inhibitory neuronal populations with recurrent coupling [7]. The dynamics of the cortical column are given by 4 coupled $2^{nd}$ order differential equations. The possible trajectories of such a system can be characterized using the phase-space representation [8]. Through simulation, these systems have been shown to have complex structures in phase-space including stationary points, limit cycles and chaotic attractors [9-11].

Perturbative activity at a stable point is often used to model spontaneous activity as seen on EEG and MEG recordings [12-14]. However, when system behavior changes dynamically over time, i.e., evinces iterant activity, it is better modelled using the full set of dynamics; involving distinct semi-stable sets, such as limit cycles and stationary points [15]. The biological manifestation of limit cycles include high amplitude oscillatory activity, physiological and abnormal activity, such as epileptic activity like seizures and spikes. We have previously established (under certain conditions) the sufficient and necessary constraints required for a neural mass system to have limit cycles [16]. In these models, the type of coupling between neuronal subpopulations determines the stability of the topology of the phase-space structure, i.e., the presence of different stable dynamic regimes. In our previous study, we showed that synaptic kernels of potential-to-current coupling can only affect the phase of the trajectories of the model, while keeping the topology constant, i.e., without any effect on the stability of limit cycles. In contrast, synaptic kernels of current-to-current coupling affect the stability of limit cycles, as does cross coupling between potential-to-current and current-to-current coupling [16]. The latter involves at least two neuronal subpopulations and allows for complex cross coupling between activity in different frequencies. Generative models with complex phase-space topology, would be well suited for analysis of empirical data, with similar complexities including, time-varying, and paroxysmal brain dynamics. This would be especially important for model inversion and identification schemes; e.g., dynamic causal modelling (DCM).

In this work, we extend the perturbative analysis presented previously to investigate the interaction between cortical columns and to derive the equations of motion for the activity in an ensemble or collection of cortical columns [16]. Interestingly, our theoretical considerations — of interacting neural mass models — allows us to derive the equations of motion for the phase of the cortical activity, which turns out to be the much-celebrated Kuramoto model [17,18]. The Kuramoto model has been used as a phenomenological model of cortical activity, but has not, to our knowledge been derived directly from more biophysical plausible models (e.g., neural mass models) [19-21]. Moreover, our derivations reveal a complex interaction between columns: where the transition between the semi-stable states of each column is determined by the (extrinsic) connectivity between columns as well as the phase difference in activity. Thus, intercortical connectivity is best described through two terms, a static (intrinsic) connectivity and a dynamic phase-dependent (extrinsic) connectivity.

Epileptic seizures have been identified as brain states with dynamically changing connectivity although a clear understanding of the underlying dynamics is outstanding. Different studies have presented conflicting results following connectivity changes during the onset, propagation and termination of a seizure [22-24]. We used our findings to construct a model of seizure propagation, allowing for inversion techniques (DCM) to estimate the connectivity matrices from EEG, recorded during epileptic seizure onset and propagation. The theoretical work presented in this paper deepens our understanding of dynamical itinerancy and provides the foundations for tractable ways of incorporating our knowledge of cortical dynamics into dynamic causal models of empirical (and clinical) electrophysiological timeseries.

# Methods

## M1. Summary of previous results

We first present a summary of the results derived in [16]. The equation of motion of a neural mass model is given by,

$$\dot{p}_i = -\omega_i^2 q_i + \varepsilon \omega_i^2 \sum g_{ij} S(q_j) + \mu \omega_i^2 \sum g_{ij} P\left(\frac{p_j}{\omega_j}\right) \qquad \text{Eq. M1.1}$$

$$\dot{q}_i = p_i$$

Where $p_i$ is the membrane current of the $i^{th}$ subpopulation of the cortical column and $q_i$ the corresponding membrane potential. $\omega_i$ is the frequency of activity of the $i^{th}$ subpopulation (or inverse half-life of the postsynaptic receptor membrane dynamics). $S$ is the synaptic kernel of potential-to-current coupling and $P$ is the synaptic kernel of current-to current coupling. $g_{ij}$ is the connection gain from the $i^{th}$ subpopulation to the $j^{th}$ subpopulation. $\varepsilon$ and $\mu$ are perturbative (coupling) constants.

This equation can be re-written using complex variables, which will simplify the derivations in the following. The complexified equation of motion is given by,

$$\dot{z}_i = -i\omega_i z_i + i\varepsilon \omega_i \sum g_{ij} S\left(\frac{z_j + z_j^*}{2}\right) + i\mu \omega_i \sum g_{ij} P\left(\frac{z_j - z_j^*}{2i}\right) \qquad \text{Eq. M1.2}$$

The complex function, $z_i$ is defined as follows,

$$z_i = q_i + i\frac{p_i}{\omega_i}$$

The complex trajectory function, $z$, can be written in modulus-argument form,

$$z_i(t) = R_i(t, \varepsilon, \mu) e^{-i(\omega_i t + \varphi_i((t, \varepsilon, \mu)))}$$

A perturbative expansion of the equations of motion in terms of $\varepsilon$ and $\mu$ obtains using the following expansion,

$$S\left(\frac{z_j + z_j^*}{2}\right) = \sum_{r=1}^{\infty} A_r \left(\frac{z_j + z_j^*}{2}\right)^{2r-1}$$

$$P\left(\frac{z_j - z_j^*}{2i}\right) = \sum_{r=1}^{\infty} B_r \left(\frac{z_j - z_j^*}{2i}\right)^{2r-1}$$

$$R_i(t, \varepsilon, \mu) = R_{i,0,0} + \sum \varepsilon^m \mu^n R_{i,m,n}(t)$$

$$\varphi_i(t, \varepsilon, \mu) = \varphi_{i,0,0} + \sum \varepsilon^m \mu^n \varphi_{i,m,n}(t)$$

Using these expansions, the left hand side (LHS) of the equation of motion (eq. M1.2) are given by,

$$LHS = \left[\sum_{k=0}^{\infty} \frac{(-i\varphi_i)^k}{k!}\right] \left(\sum \varepsilon^n \mu^m \dot{R}_{i,n,m}(t)\right) e^{-i\omega_i t} - i\left(R_{i,0,0} + \sum \varepsilon^n \mu^m R_{i,n,m}(t)\right)$$
$$* \left(\sum \varepsilon^n \mu^m \dot{\varphi}_{i,n,m}(t)\right) \left[\sum_{k=0}^{\infty} \frac{(-i\varphi_i)^k}{k!}\right] e^{-i\omega_i t}$$

The right-hand side (RHS) of the equation of motion ( eq. M1.2) is given by,

$$RHS = i\omega_i \varepsilon \sum g_{ij} \sum_{r=1}^{\infty} A_r \frac{\left(R_{j,0,0} + \sum \varepsilon^n \mu^m R_{j,n,m}(t)\right)^{2r-1}}{2^{2r-1}}$$
$$* \left[\sum_{l=0}^{2r-1} \binom{2r-1}{l} e^{-i((2r-1)-2l)(\omega_j t)} \left[\sum_{k=0}^{\infty} \frac{(-i(2r-1-2l)\varphi_j)^k}{k!}\right]\right]$$
$$+ i\omega_i \mu \sum g_{ij} \sum_{r=1}^{\infty} (-1)^r B_r \frac{\left(R_{j,0,0} + \sum \mu^n R_{j,0,n}(t)\right)^{2r-1}}{2^{2r-1}}$$
$$* \left[\sum_{l=0}^{2r-1} (-1)^l \binom{2r-1}{l} e^{-i((2r-1)-2l)(\omega_j t)} \left[\sum_{k=0}^{\infty} \frac{(-i(2r-1-2l)\varphi_j)^k}{k!}\right]\right]$$

An adiabatic approximation — over a complete cycle of oscillation of the cortical column — can now be derived: The changes in the trajectory after integrating over a full cycle — defined as the derivative of the flow in the new time scale — revealed that the modulus (or amplitude variable) was only affected by *P*-coupling (first order expansion terms) and combinations of *P*- and *S*-coupling (2nd order expansion terms),

$$\frac{dR_{i,0,1}}{dt} = \omega_i g_{ii} \sum_{r=1}^{\infty} B_r \frac{R_{i,0,0}^{2r-1}}{2^{2r-1}} \binom{2r-1}{r-1} \qquad \text{Eq. M1.3}$$

$$\frac{dR_{i,1,1}}{dt} = \sum \frac{g_{ij} g_{ji} \omega_i \omega_j}{\omega_j^2 - \omega_i^2}$$
$$* \sum_{r,s=1}^{\infty} [(r-1)A_r B_s \omega_j + r A_s B_r \omega_i] \frac{R_{j,0}^{2r-2} R_{i,0,0}^{2s-1}}{2^{2r+2s-3}} \binom{2r-1}{r}\binom{2s-1}{s} \qquad \text{Eq. M1.4}$$

The ensuing phase-space structure depends on the changes in modulus (i.e., amplitude) and not the argument variable ($\varphi$, or equivalently the phase variable). The stable points (equivalently the roots of the equations on the RHS) define the limit cycles of the original equations describing the cortical column. This is determined by the functional form of the synaptic kernels, *S* and *P.* Note that diagonal terms of the connectivity matrix, *g*, determine eq M1.3 and the off-diagonal terms determine eq. M1.4. Adding Eq M1.3 and M1.4 furnishes the resultant flow of the amplitude dynamics of a cortical column up to 2nd order in the perturbation expansion. Using this equation, the presence of limit cycles can be determined for a single cortical column (we have previously presented results up to 1st order), [16].

**M2. Interaction between off-diagonal and diagonal terms within a cortical column**

In this section, we investigate and contrast the effect of self-coupling (represented by diagonal terms of the connectivity matrix) and coupling between different subpopulations within a cortical column (represented by off diagonal terms of the connectivity matrix). We will show that the presence of limit cycles and their stability is determined by the connectivity matrix and the form of the synaptic kernels.

The equation of motion of the amplitude variable (modulus variable) is given by,

$$\frac{dR_i}{dt} = \mu\omega_i g_{ii} \sum_{r=1}^{\infty} B_r \frac{R_{i,0}^{2r-1}}{2^{2r-1}} \binom{2r-1}{r}$$
$$+ \mu\varepsilon \sum \frac{g_{ij}g_{ji}\omega_i\omega_j}{\omega_j^2 - \omega_i^2}$$
$$* \sum_{r,s=1}^{\infty} [(r-1)A_r B_s \omega_j + r A_s B_r \omega_i] \frac{R_{j,0}^{2r-2} R_{i,0,0}^{2s-1}}{2^{2r+2s-3}} \binom{2r-1}{r}\binom{2s-1}{s}$$

In a 2 neural subpopulation system, the stable points of the diagonal components (self-interaction terms) are stable for small off-diagonal components (intra-cortical connections between different neuronal subpopulations) as shown in Figure 1. This can also be shown, in more general terms, by estimating the eigenvalues of the Jacobian of the flow at the stationary points.

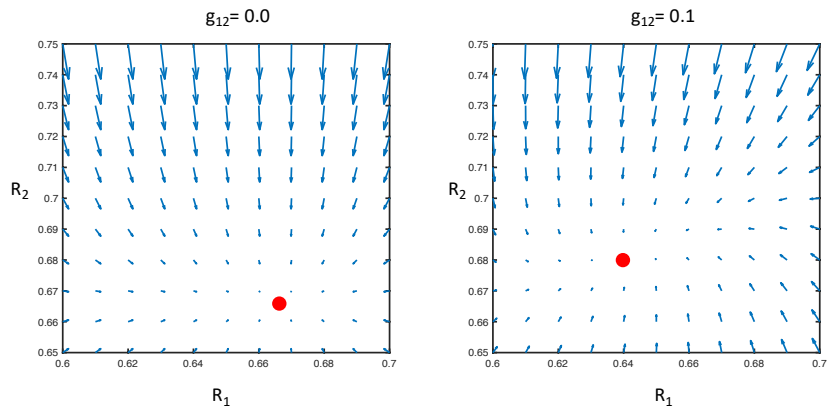

**Figure 1. Interaction between 2 neuronal subpopulations is demonstrated, where the amplitude variables for 2 populations are shown ($R_1$ and $R_2$). The vector plot around the stable point for different off-diagonal couplings is shown with no off-diagonal terms to the left and small off-diagonal terms to the right. Small off**

**diagonal terms result in a small deviation of the flow but retains the stability of the stationary points, which represents limit cycles in the original model. The red ball indicates a stable stationary point.**

If the off-diagonal components are small in comparison to the diagonal terms there will be $r^N$ stable states, r for each of the N populations in a column. The integer r is determined by the synaptic kernel *P*. However, if this condition is not fulfilled (i.e., with larger off-diagonal terms) the dynamics will change considerably. For a 2 subpopulation cortical column the number of limit cycles and their stability will change, as illustrated in Figure 2.

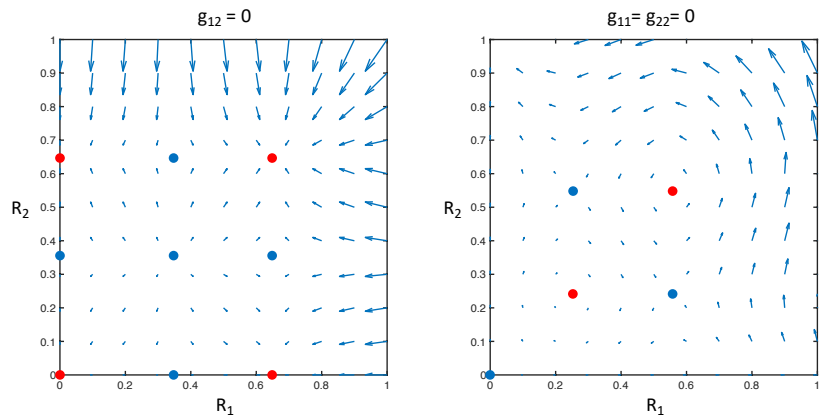

**Figure 2. Vector plot describing phase-space dynamics for different connectivity matrices. The panel to the left shows 4 stable (in red) and 5 unstable stationary points (in blue) when the off-diagonal components are 0. All of these stationary points would represent stable or unstable limit cycles of the original model. The panel to the right shows the dynamics with only off-diagonal terms in the connectivity matrix. 2 stable (in red) and 3 unstable stationary points (in blue) are seen representing limit cycles of the original model.**

The trajectories for off-diagonal coupling within a cortical column show chaotic itinerancy, as illustrated in figure 3. This itinerancy increases as the number of limit cycles per cortical column increases — and the trajectories start folding tightly around each other.

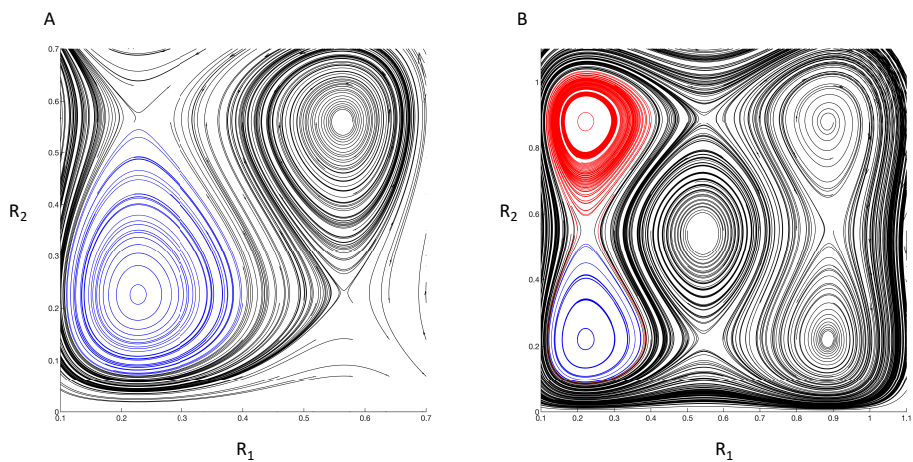

**Figure 3. Trajectories are plotted in phase-space for off-diagonal coupling between two sub-populations within a cortical column. A) The *P*-coupling has 3 zero points resulting in one stable point (representing 1 stable limit cycles in the phase-space representation) with attracted trajectories drawn in blue. B) The *P*-coupling has 5 zero points resulting in two stable points with attracted trajectories drawn in blue and red.**

**Note the complicated interaction between the trajectories for the two stable points with the attractor of one stable point (in blue) placed within the attractor of the other (in red). These points, again, represent stable limit cycles in the phase space representation.**

# Results

In this section, we characterize the interaction between cortical columns; estimating the dynamics of each column induced by the recurrent (extrinsic) coupling between them. We will further show that the interaction between cortical columns is not only determined by the coupling between them but also their phase difference (see section R1 for details). We then study the phase interaction between columns, showing that the equations of motion reduce to the Kuramoto model (see section R2 for details). Using the results from R1 and R2, we then derive the stability of semi-stable states, i.e., their transition times (see section R3 for details). Having established the functional form of the model, we then consider robust methods for model inversion; namely, estimating the connectivity matrix from transition times (section R4). Finally, all the results from R1-4 are combined to derive a scheme for estimation of intercortical connectivity from seizure propagation, in section R5.

## R1. Interaction between 2 cortical columns

The dynamics of a cortical column is easily tractable when including only the diagonal terms as shown in the Methods section. Moreover, the diagonal terms (1$^{st}$ order terms in the perturbation analysis) are larger in amplitude than the off-diagonal terms (2$^{nd}$ order terms in the perturbative analysis). For this reason, the interaction between cortical columns was investigated here, with only diagonal entries in the connectivity matrix, i.e., only using the *P*-coupling term and neglecting the P to S coupling terms. To further simplify the analysis, we will assume only 1 diagonal entry, although the derivations could be expanded — with some increase in computational complexity — with several diagonal terms.

We first introduce 2 sets of couplings between the subpopulations in the 2 cortical columns. We have so far studied the dynamics induced by intracortical (or intrinsic) connectivity which will be denoted by $g_{ij}$. The intercortical (or extrinsic) connectivity will be denoted by $h_{ij}$. We then have a new set of equations determining the motion of the 2 cortical columns.

$$\left( \dot{R}_i^1(t,\varepsilon,\mu) e^{-i(\omega_i t + \varphi_i^1)} - i R_i^1(t,\varepsilon,\mu) \dot{\varphi}_i^1 e^{-i(\omega_i t + \varphi_i^1)} \right)$$
$$= i\varepsilon\omega_i \sum g_{ij} S_j^1 + i\mu\omega_i \sum g_{ij} P_j^1 + i\varepsilon\omega_i \sum h_{ij} S_j^2 + i\mu\omega_i \sum h_{ij} P_j^2$$

Eq. R1.1

As was summarized in the method section, the first term on the RHS will not have any effect on the phase space structure up to 2$^{nd}$ order. The second term affects the phase-space structure, due to self-connection terms within a given column. The 3$^{rd}$ and 4$^{th}$ terms will also affect the phase-space structure, due to phase differences between identical sub-populations in the two cortical columns. The expansion for *R* will not change but that for the phase variable (to include phase differences) is given by,

$$\varphi_i^1(t,\varepsilon,\mu) = \varphi_{i,0,0}^1 + \sum \varepsilon^m \mu^n \varphi_{i,m,n}^1$$

Different cortical columns are indexed by superscripts. We now examine the effect of the intercortical connectivity matrix and start with the 3$^{rd}$ term in eq R1, the *S*-coupling term. Keeping only 1$^{st}$ order terms gives, after a few manipulations,

$$\frac{dR^1_{i,1,0}}{dt} - iR^1_{i,0,0}\frac{d\varphi^1_{i,1,0}}{dt} = i\omega_i h_{ii} \sum_{r=1}^{\infty} A_r \frac{{R^2_{i,0,0}}^{2r-1}}{2^{2r-1}} \binom{2r-1}{r-1} e^{i(\varphi^1_{i,0,0}-\varphi^2_{i,0,0})}$$

As was shown previously, there is also a change in the phase variable due to self-connection terms within a cortical column,

$$iR^1_{i,0,0}\frac{d\varphi^1_{i,1,0}}{dt} = -i\omega_i g_{ii} \sum_{r=1}^{\infty} A_r \frac{{R^1_{i,0,0}}^{2r-1}}{2^{2r-1}} \binom{2r-1}{r-1}$$

We then get the following for the dynamics with coupling between the phase and amplitude variables,

$$\frac{dR^1_{i,1,0}}{dt} - iR^1_{i,0,0}\frac{d\varphi^1_{i,1,0}}{dt}$$
$$= i\omega_i g_{ii} \sum_{r=1}^{\infty} A_r \frac{{R^1_{i,0,0}}^{2r-1}}{2^{2r-1}} \binom{2r-1}{r-1}$$
$$+ i\omega_i h_{ii} \sum_{r=1}^{\infty} A_r \frac{{R^2_{i,0,0}}^{2r-1}}{2^{2r-1}} \binom{2r-1}{r-1} e^{i(\varphi^1_{i,0,0}-\varphi^2_{i,0,0})}$$

The *S*-coupling term as a first order contribution to both the phase and amplitude dynamics. In a similar way, the 2$^{nd}$ and 4$^{th}$ term in eq. R1 will also have a first order contribution (from connectivity terms that couple populations with the same frequency of oscillation, $g_{ii}$ and $h_{ii}$). Using previous results from the Method section, one can derive the following.

$$\frac{d}{dt}R^1_{i,0,1} - iR^1_{i,0,0}\frac{d}{dt}\varphi^1_{i,0,1}$$
$$= \omega_i g_{ii} \sum_{r=1}^{\infty} B_r \frac{{R^1_{i,0,0}}^{2r-1}}{2^{2r-1}} \binom{2r-1}{r-1}$$
$$+ \omega_i h_{ii} \sum_{r=1}^{\infty} B_r \frac{{R^2_{i,0,0}}^{2r-1}}{2^{2r-1}} \binom{2r-1}{r-1} e^{i(\varphi^1_{i,0,0}-\varphi^2_{i,0,0})}$$

$$\frac{d}{dt}R^1_{i,0,1} = \omega_i g_{ii} \sum_{r=1}^{\infty} B_r \frac{{R^1_{i,0,0}}^{2r-1}}{2^{2r-1}} \binom{2r-1}{r-1} + \omega_i h_{ii} \cos(\varphi^1_{i,0,0} \quad\quad \text{Eq. R1.2}$$
$$- \varphi^2_{i,0,0}) \sum_{r=1}^{\infty} B_r \frac{{R^2_{i,0,0}}^{2r-1}}{2^{2r-1}} \binom{2r-1}{r-1}$$

$$\frac{d}{dt}\varphi^1_{i,0,1} = -\omega_i h_{ii} \sin(\varphi^1_{i,0,0} - \varphi^2_{i,0,0}) \sum_{r=1}^{\infty} B_r \frac{R^{2\,2r-1}_{i,0,0}}{R^1_{i,0,0}\, 2^{2r-1}} \binom{2r-1}{r-1}$$ Eq. R1.3

A similar relation will hold for $R_i^2$ and $\varphi_i^2$. As discussed previously, the semi-stable states of one cortical column (stationary point at 0 and a stable limit cycle with amplitude $R_s$) with one self-connecting loop (one *P-term*) are given by,

$$\{0\}, \{R_s\}$$

Two interconnected columns will have the following stable states for a weak intercolumnar connection i.e., $|g_{ii}| \gg |h_{ii}|$

$$\{0,0\}, \{0, R_s\}, \{R_s, 0\}, \{R_s, R_s\}$$

Phase portraits can be constructed for different values of the phase lag (φ) between the cortical columns. The phase portraits have the same topological structure but distinct dynamics. Namely, there is a change in the size of the attractors of the stable states: see Figure 4.

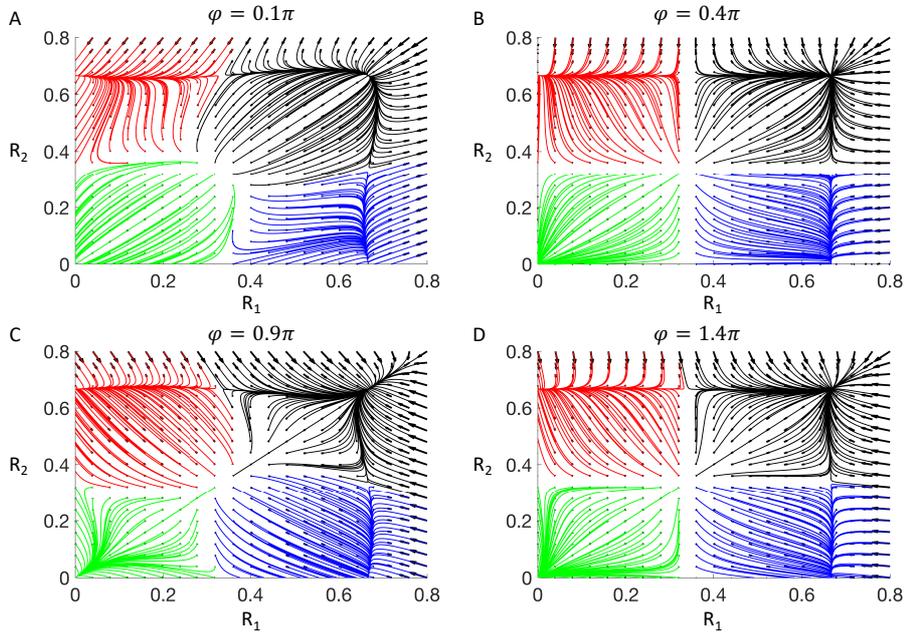

**Figure 4. Phase-space plots with trajectories for different values of phase lag (φ). There are 4 stable states and the trajectories that are attracted to each stable point are drawn with a different color. Each stable point represents a limit cycle in the dynamics of the cortical column. The attractors for the stable points can be seen to depend on phase lag.**

We now focus on the state transition:

$$\{0, R_s\} \rightarrow \{R_s, R_s\}$$

See appendix (S1) for a detailed derivation of the results. The equation of motion for the trajectories between the two stable states can be approximated by the following ($A_1$ is a constant),

$$\frac{d}{dt} R^1_{i,0,1} = \omega_i g_{ii} \sum_{r=1}^{\infty} B_r \frac{{R^1_{i,0,0}}^{2r-1}}{2^{2r-1}} \binom{2r-1}{r-1} + \omega_i h_{ii} \cos(\varphi^1_{i,0,0} - \varphi^2_{i,0,0}) A_1 \qquad \text{Eq. R1.4}$$

Integrating this equation along $R_1$ gives us the potential determining the trajectories near the stable points.

$$U = -\omega_i g_{ii} \sum_{r=1}^{\infty} B_r \frac{{R^1_{i,0,0}}^{2r}}{r 2^{2r}} \binom{2r-1}{r-1} - \omega_i h_{ii} \cos(\varphi^1_{i,0,0} - \varphi^2_{i,0,0}) A_1 R^1_{i,0,0} \qquad \text{Eq. R1.5}$$

The potential depends on the phase lag between the cortical columns and noise process affecting the columns: see Figure 5. This analysis shows that phase differences close to 0 have a global stable point at $R_s$, while phase differences close to $\pi$ have a global stable point at 0.

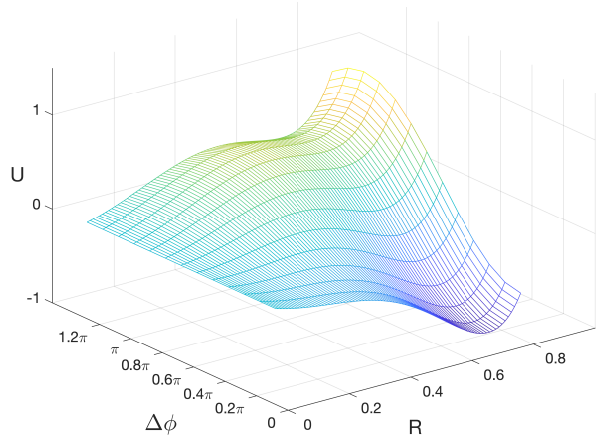

**Figure 5. Trapping potential for different values of phase lag (φ). With 0 phase lag the stable point at $R_s$ has global stability while at phase lag $\pi$ the global stability is at R=0.**

The transition rates between semi stable states for a given network of cortical columns depend on the phase lag between the columns (and the coupling). The transition rates define how the state of one cortical column affects the state of a second cortical column. The transition rate can also be seen as a connectivity measure between the two columns as it defines the stability. We will use this measure to define a dynamic connectivity between cortical columns — a measure that, crucially, can be estimated from empirical data, as will be shown below.

This dynamic connectivity could play a substantive role in seizure propagation, and the dynamic nature of connectivity indicated by experimental data; where various connectivity measures between different cortical regions vary over time, especially during seizure progression. On the basis of the current analysis, one might propose that the dynamic variation in connectivity — seen in empirical data — is at least partly dependent on the phase of ongoing neuronal oscillations. This proposal will be familiar to readers versed in hypotheses such as communication through coherence and related formulations of dynamic connectivity [1-9].

**R2. Phase dynamics**
The phase lag and amplitude dynamics are coupled; however, as an approximation (mainly to keep the results tractable) we uncouple them by using the average amplitude values (averaged over different trajectories).

$$R_i^k \to \langle R_i^k \rangle$$

We then get the following,

$$\frac{d}{dt}\varphi_{i,0,1}^1 = -\omega_i h_{ii} \sin(\varphi_{i,0,0}^1 - \varphi_{i,0,0}^2) \sum_{r=1}^{\infty} B_r \frac{\langle R_i^2 \rangle^{2r-1}}{\langle R_i^1 \rangle 2^{2r-1}} \binom{2r-1}{r-1} \quad \text{Eq. R1.5}$$

This equation approximates the activity around the stationary sets (stationary point or limit cycles) of the cortical columns. It can be seen that it is equivalent to the celebrated Kuramoto model and can also be derived from the Hamiltonian of the XY-model [25]. This model has been used to describe phase transitions in solid-state physics and has been shown to have interesting topological phase transitions as was shown in [26]. Systems close to local equilibrium states — as approximated in equation eq R1.5 – have excitations with energies approaching 0; indicating long range correlations at low temperatures (goldstone modes or bosons), which were first described in the analysis of superconductivity by Nambu and Goldstone [27-28]. The presence of low energy excitations allows for spatial correlations to take place over large spatial domains in the cortex, which could be of importance in the collation of data in the human brain

**R3. Simulations of state transitions**
The above derivations indicate that state transitions depend on the phase difference between columns. To further characterize this dependency, 2 cortical columns were interconnected and simulated (using the Euler-method and neural mass equations). The effects of intercolumnar connectivity and phase difference were studied using this numerical analysis. One column was allowed to oscillate at a limit cycle and did not receive feedback from the second. The phase lag between the two cortical columns was fixed, to estimate the effect of connectivity coupling and phase lag on the dynamics of the second cortical column.

With fixed connectivity between the columns, the transition rate between the two metastable states of the second column was determined for different phase lags. The results indicate the existence of a dynamic and static connectivity where the transition rate is determined by the dynamic connectivity. This can be switched on or off, relatively quickly, depending on the phase lag between the interacting columns: see Figure 6. Simulations and theoretical work indicate that larger networks of interacting systems produce a turbulent flow (in 2 dimensions) of phases, where discontinuities are created by vortices [21,25,26]. The presence of similar vortices on cortical networks could cause rapid, perhaps turbulent, changes in phase lags, which in turn would cause changes in the "dynamic" connectivity. This could alter the propagation pathways of cortical activity in an itinerant fashion.

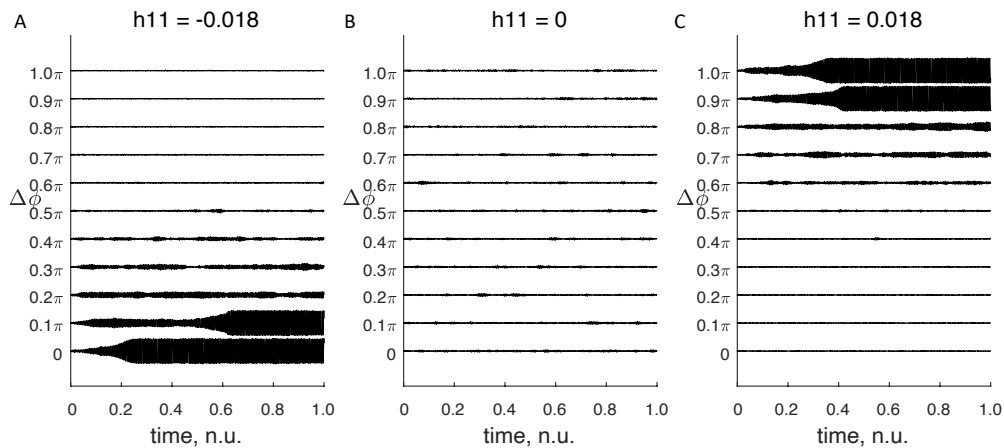

**Figure 6. A. Activity of a cortical column (with a meta-stable state at the origin and a limit cycle) when under the influence of a second cortical column oscillating in the limit cycle state. The activity — for small phase lags – transitions to limit cycle oscillations (with an increase in amplitude of the oscillation as seen in the graphs). The transition rates depend on the phase lag between the columns. B. With no connection between the columns ($h_{11}$=0) there will be no transition to the limit cycle oscillation during the 1 time unit plotted. C. Similar activity as A but the connection between the cortical columns is "switched on" for phase lags corresponding to $\pi$ and is "switched off" for phase lags around 0. Note the opposite sign of intercolumnar connection for A and C ($h_{11}$).**

## R4. Inverse modelling to determine synaptic intercolumnar connectivity

We now construct a forward model, which maps the connectivity matrix to propagation times for the transition from low amplitude activity to high amplitude oscillations (i.e., a proxy for seizure activity) for different cortical columns. The output will be the estimated time for the dynamics of a cortical column to transit between the meta-stable states. The potential driving the transition between the 2 stable states was estimated in section R1 (eq. R1.5).

$$U = -\omega_i g_{ii} \sum_{r=1}^{\infty} B_r \frac{{R^1_{i,0,0}}^{2r}}{r 2^{2r}} \binom{2r-1}{r-1} - \omega_i h_{ii} \cos(\varphi^1_{i,0,0} - \varphi^2_{i,0,0}) A_1 R^1_{i,0,0}$$

The Fokker Planck equation resulting from the flow of the above gradient is given by (amplitude of the noise process will be denoted by $\sigma_n$)

$$\frac{\partial \rho}{\partial t} = -\frac{\partial}{\partial x}\left(\rho \left(\frac{\partial}{\partial x} U\right)\right) + \frac{\sigma_n^2}{2} \frac{\partial^2 \rho}{\partial x^2} = L_{FP}(\rho) \qquad \text{Eq. R4.1}$$

$$\partial \rho = L_{FP}(\rho) \partial t$$

This Fokker Planck flow will be projected to the space of normal distributions. The flow of the distributions can then be parameterized using the mean (μ) and the standard deviation (σ) of the true distribution.

We will assume that $h_{ii} > g_{ii}$ which will give is the following,

$$\frac{d\mu}{dt} = -\omega_i h_{ii} \cos(\varphi^1_{i,0,0} - \varphi^2_{i,0,0}) A_1 \propto -h_{ii} \cos(\varphi^1_{i,0,0} - \varphi^2_{i,0,0})$$

Which in turn can be integrated into the following expression for the mean (μ) parameter,

$$\mu(t) = \mu(0) - h_{ii} \cos(\varphi^1_{i,0,0} - \varphi^2_{i,0,0}) t$$

This indicates that the transition rate between the low and high amplitude state is directly proportional to the value of *h* and depends on the phase lag between the two columns (as was shown in the simulation in R4) and will only occur if the drift of the mean is positive (i.e., a drift from low amplitude to large amplitude activity), Figure 7. Note that when the cosine of the phase lag is near zero, the true transition time will be given by terms related to $g_{ii}$, see Appendix S2 for the full expression for the transition rate for the mean. The transition time is defined as the time for the mean of the distribution to drift from the stable state at 0 to the attractor of the limit cycle.

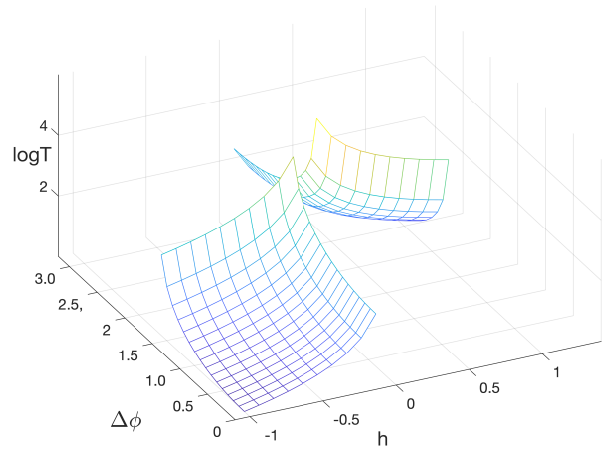

**Figure 7. The logarithm of the transition time from the low amplitude dynamics to the high amplitude dynamics of a cortical column under the influence of a second cortical column. The static connectivity between the columns is given by *h* and the phase lag by Δφ.**

### R5. Modelling seizure propagation

Each cortical column has two stationary states of activity, a stable point around the origin and a stable limit cycle. On this basis, one can explore the spread of high amplitude activity (as a proxy for electrographic seizure activity) through a network of cortical columns. The dynamics of a given cortical column is given by,

$$\frac{d}{dt}R_i = \omega_i g_{ii} \sum_{r=1}^{\infty} B_r \frac{R_i^{2r-1}}{2^{2r-1}} \binom{2r-1}{r-1} + \sum_j \omega h_{ji} \cos(\varphi_i - \varphi_j) A_j \qquad \text{Eq. R5.1}$$

We want to model seizure onset and the connections of the network will be assumed to be cascading, i.e., areas involved later in the seizure do not connect to areas involved earlier: see Figure 8. Relaxing this assumption would allow for the study of seizure termination, however, the complexities of the model increases, and we will not pursue that further in this paper.

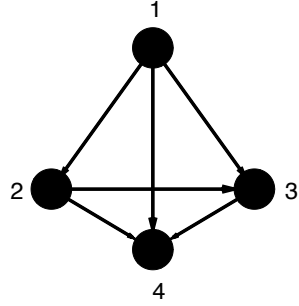

**Figure 8.** Network with 4 nodes simulating seizure progression. Connections are unidirectional from nodes involved earlier to nodes involved later in the seizure propagation.

As a typical example, we will simulate seizure onset times ($T_j$) for N electrodes with an unknown connectivity matrix, which will be estimated using the onset times and phase lag values between each electrode pair. The mean of the distribution of the $i^{th}$ population can be approximated (after absorbing constants into the definition of *h*) using the following equation ($\Theta$ is the Heaviside step function).

$$\mu_i(t) = \mu_i(0) - \sum_{j<i}^{N} h_{ji} \cos(\varphi_i - \varphi_j) \int_0^t dt \Theta_j(t' - T_j)$$

$$\Theta_j(t' - T_j) \approx \frac{1}{2} + \frac{1}{2} \tanh(t' - T_j)$$

$$\int_0^t dt \frac{1}{2} + \frac{1}{2} \tanh(t' - T_j) = \frac{t}{2} + \frac{1}{2} \ln(\cosh(t' - T_j)) - \frac{1}{2} \ln(\cosh(T_j))$$

$$\mu_i(t) = \mu_i(0) - \sum_{j<i}^{N} h_{ji} \cos(\varphi_i - \varphi_j) \left[ \frac{t}{2} + \frac{1}{2} \ln(\cosh(t - T_j)) - \frac{1}{2} \ln(\cosh(T_j)) \right]$$

We take $\mu_i(0) = 0$, so that the transition time is defined by the following equation.

$$0 = 1 + \sum_{j<i}^{N} h_{ji} \cos(\varphi_i - \varphi_j) \left[\frac{T_i}{2} + \frac{1}{2}\ln(\cosh(T_i - T_j)) - \frac{1}{2}\ln(\cosh(T_j))\right] = F_i(\vec{T})$$

The solution to the following equation provides a prediction of (measurable) onset times as a function of (unmeasurable) connectivity, **h**.

Eq. R5.2
$$\vec{F}(\vec{T}) = 0$$

In other words, this equation provides a generative or forward model relating latent dynamic connectivity to observable onset times. The onset times are determined by the connectivity matrix, **h**, and phase lag between the cortical columns. The *A-term* (Eq. R5.1) will have a large effect on the dynamics when the activity of the *j*[th] column is in a limit cycle. To investigate the validity of these derivations numerically, seizure spread was simulated using the neural mass model for interacting cortical columns above. The Euler method and an underlying noise process was used for the simulation. It can be seen that seizure progression is not unique for a given connectivity matrix as this is determined by a mixture of deterministic and stochastic variables: see Figure 9.

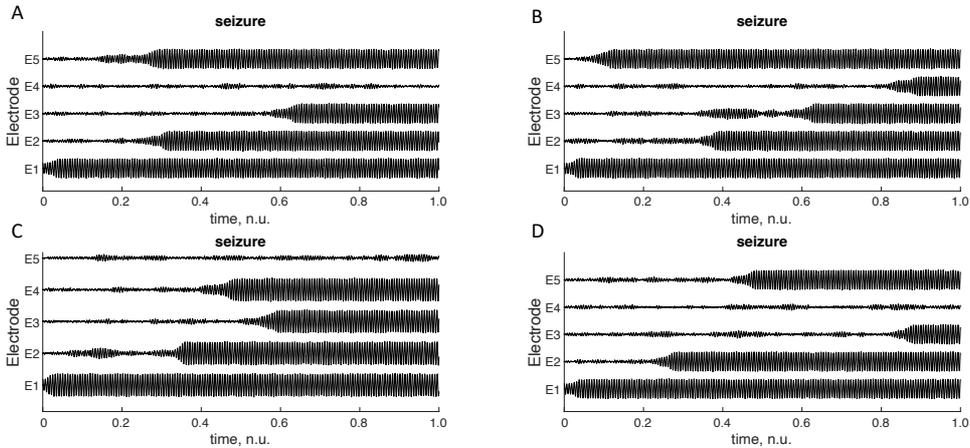

**Figure 9. Seizure propagation simulated using 5 interconnected nodes or columns. Note that seizure propagation varies for each trial even though the connectivity matrix is kept constant, illustrating the stochasticity of the propagation.**

The distribution of seizure onset for the same connectivity matrix was estimated using 1000 simulations of the same model, using fitted normative distributions: see Figure 10. Using only the summary statistics (mean and standard deviation) of seizure onset times — across multiple realizations — we inverted our model for seizure propagation (using DCM) (eq. R6.2). This inversion uses standard variational procedures to recover an estimate of the connectivity matrix from the statistics of the simulated onset times. See Figure 11 for details of this inversion. In general, the estimated connections (posterior estimates) were nearer the baseline values — used to generate the synthetic data — than the prior values. However, the change between the prior and the posterior distribution is determined by the precision of the prior, and also subject to the risk of estimating posteriors at local minima — factors that can be nuanced to the specific hypothesis being tested [29]. This analysis is presented as proof of principle that differences in extrinsic synaptic connectivity among cortical columns is expressed in the (statistics of) seizure onset times at the respective sources. And is expressed in a way that enables estimation of differences in connectivity by careful modelling of these measures.

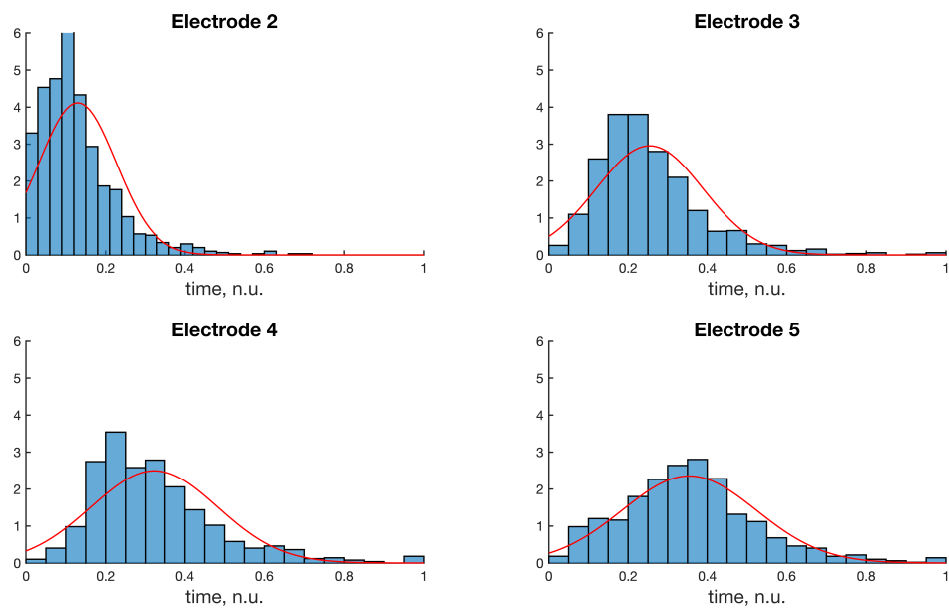

**Figure 10. Probability distributions for seizure onset for different electrodes (or cortical columns) with fitted probability distributions for the mean.**

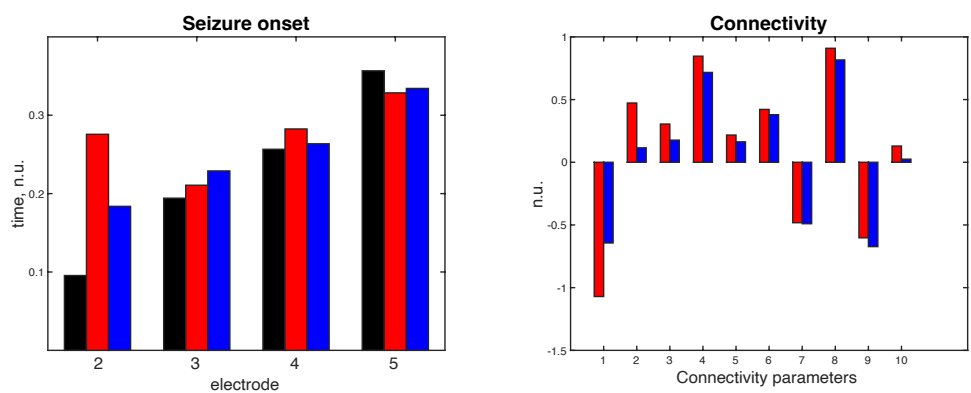

**Figure 11. Seizure onset times are given for the baseline data (black), estimated onset times for the data simulated using the prior (red) and posterior estimation (blue) of the connectivity matrix. A comparison between the prior (red) and posterior (blue) connectivity estimates is given on the right, where the baseline values were normalized to 0. In other words, the connectivity in the right panel is the difference between the**

**estimates and baseline connectivity used to generate the synthetic data. The posterior estimates of the connectivity values tend towards the baseline values (blue columns are in general smaller in amplitude than the red columns).**

## Discussion

In this study, we continued our theoretical investigation into the dynamics of cortical columns; in particular, the interaction among columns. We have previously established the sufficient and necessary criteria for meta-stable states within a cortical column. Metastability was shown to depend on the synaptic kernels connecting neuronal subpopulations within a column. The topological structure of the phase portrait was determined by synaptic kernels modelling current-to-current coupling. Moreover, it was shown that only those connections between neural subpopulations which led to a "resonance" of activity survived the adiabatic averaging and could affect the dynamics. For a single cortical column resonating connections were underwritten by self-connecting loops, involving one or more neural subpopulations. These meta-stable states within a cortical column remained stable, at least, for week coupling between cortical columns.

Even though intracortical (intrinsic) connectivity required self-connecting loops to give resonating terms, a mixture of intercortical (extrinsic) and intrinsic connectivity resulted in resonating terms from open ended connections (i.e., a connection between different neuronal subpopulations). This would manifest if neural subpopulations from different cortical columns oscillating at the same frequency were connected (directly or through other types of neuronal subpopulations). The presence of open ended connections introduced a new variable into the dynamics of cortical columns; the phase lag between homologous neural subpopulations in different columns. The ensuing dynamics involve coupled equations of amplitude and phase variables. From these equations of motion, we can approximate the rate of transitions between meta-stable states in networks of connected columns — showing that the transition rate depends on extrinsic connectivity and also the phase difference between columns.

We propose that the transition rate reflects a dynamic connectivity measure that is phase dependent and can change quickly: in fact, the possibility exists that any connection can be turned on and off as the dynamical system evolves. There is plenty of support in empirical study of cortical activity, measured with micro or macro electrodes, for this kind of dynamic connectivity. This is often seen in highly non-linear or turbulent activity of the cortex for example, the progression of epileptic seizures [22-24]. Investigating the equations of motion revealed that the phase dynamics, at least when perturbed around a meta-stable state, is given by the XY-model (or equivalently the Kuramoto model). Similar proposals of quickly changing connectivity have been presented when analyzing brain activity using the Kuramoto model [20]. Furthermore, there is a vast literature on the dynamics of the XY model showing that there is a topological phase transition; where the continuous symmetry of the XY-model (or the phase-dynamics presented this study) is broken. The implications for columnar network dynamics would include long range order for the phase variables at low temperatures, which collapses at a critical temperature, resulting in an exponential decay in phase correlations. This transition has been explained by a change in the topological structure of the dynamics at a critical temperature where low temperature dynamics is governed by stable vortex-antivortex pairs and high temperature activity by free vortices [25].

In contrast to the majority of studies — applying the Kuramoto model to cortical activity — we have derived it as an emergent property of a network of neural mass models. This strengthens the assumptions of our derivations and, furthermore, yields the coupled amplitude dynamics of cortical activity that is usually not associated with the Kuramoto model. The coupling between phase and amplitude dynamics is what allowed us to define *dynamic connectivity*. This was used as a generative model (i.e., a forward model) of seizure propagation within a network of cortical columns. This model was compared to simulated neural mass model activity, which gave similar findings. Furthermore, we inverted this model using simulated data to estimate the unknown extrinsic connectivity. The results showed an improved connectivity estimate, approaching the values used to simulate the data; however, these estimates depend on the prior values and other issues related to the identifiability of this sort of model — issues that can be nuanced depending on the specific question asked [29].

Here, we just present the model inversion as a proof of principle that patterns of extrinsic connectivity can be recovered — to a greater or lesser degree — under our seizure progression model. We envisaged this model could be further expanded with multifrequency dynamics, which is often seen in clinical data on seizure progression, by including more meta-stable states per cortical column. One of the main clinical implications of this computational exercise is the provision of a generative model for seizure progression that is accompanied by a relatively straightforward inversion scheme. In principle, one could estimate a connectivity matrix from EEG data from seizure progression with a model which 1) specifically models seizure progression 2) is based on the XY-model (equivalently the Kuramoto model) derived from basic neural mass models and 3) biophysically linked to synaptic connectivity. An important clinical question — which arises for any patient where epilepsy surgery is considered — is to define the optimal region of the brain which, when removed would result in seizure freedom. This can now be modelled prior to surgery by removing or disconnecting parts of the cortical network and estimating the effect that has on seizure propagation. This is clearly speculative but could be used retrospectively on a surgical case series to investigate the possible usefulness of this model and also the modifications required to improve the model.

We have not investigated seizure termination in this study, which is of clinical interest as it will define the duration of seizures and hence the clinical impact the seizures will have on a patient. The model we proposed for seizure propagation is a "cascading" network with nonrecurrent connectivity, which can model the propagation of seizure activity but not its termination. Using a recurrent connectivity matrix, it is possible to model seizure termination, but results in a forward model of greater computational complexity but could remain tractable for model inversion. This will be further investigated in a future study of seizure propagation and termination.

The equations of motion revealed that cortical dynamics is only related to phase differences and not the absolute value of phases. The corresponding Hamiltonian for the system would also only depend on phase differences between columns. This is in contrast to amplitude values, which populate freely the equations of motion. The property of phase differences in the Hamiltonian results in excitation levels for the system around a semi-stable equilibrium state of infinitely small energies, i.e., very slowly varying fields [27,28]. These (Goldstone) modes result in long range correlations between cortical columns. They contain information about the correlation between a large number of cortical columns. If it is assumed that cortical

columns — in the healthy brain – project information onto semi-stable states we can now define the process by which the total information represented in the brain can be summarized. In other words, Goldstone modes might be projections of the information processed by the brain; moreover, they are also known to cause the turbulent flow in phase dynamics — something we have suggested might be a key feature in seizure dynamics. One could entertain the hypothesis that cognitive function and seizure threshold are both dependent on goldstone modes. This would imply that there is a bound on cognitive function, constructed using cortical columns, as the disruption of turbulent phase dynamics is dependent on the same Goldstone modes. Interestingly, it is probable that epileptic seizures affect a brain only if it has a minimal complexity; seizures are only known in higher order animals (mammals) [30]. We will — in a future study — investigate the implications that these low energy excitations have on the dynamics of interconnected cortical columns.

In conclusion, we have derived the dynamics of interconnected cortical columns, where the phase and amplitude were shown to be the definitive variables. The equations of motion revealed that excitations of activity around stationary states were governed by the Kuramoto model — a model that has often been used to model phase dynamics in the brain. The results that were derived were used to construct a forward model for seizure propagation, which was inverted using DCM of simulated data. In future work, we will apply this model to actual seizure data investigating the possible clinical impact this might have. Moreover, we hope to examine the above theory of cortical connectivity from the perspective of a cortical lattice theory and investigate the implications for modelling cortical function.

**Funding statement**

GC was supported by funding from Karolinska University Hospital. KJF is supported by funding for the Wellcome Centre for Human Neuroimaging (Ref: 205103/Z/16/Z) and a Canada-UK Artificial Intelligence Initiative (Ref: ES/T01279X/1). RER is supported by funding from the Wellcome Trust (209164/Z/17/Z). KJF and RER are supported by the European Union's Horizon 2020 Framework Programme for Research and Innovation under the Specific Grant Agreement No. 945539 (Human Brain Project SGA3).

# S. Appendix

## S1. Transition time between meta-stable states for 2 connected cortical columns

The equation governing the flow of trajectories is given by,

$$\frac{d}{dt}R^1_{i,0,1} = \omega_i g_{ii} \sum_{r=1}^{\infty} B_r \frac{{R^1_{i,0,0}}^{2r-1}}{2^{2r-1}} \binom{2r-1}{r-1} + \omega_i h_{ii} \cos(\varphi^1_{i,0,0}$$
$$- \varphi^2_{i,0,0}) \sum_{r=1}^{\infty} B_r \frac{{R^2_{i,0,0}}^{2r-1}}{2^{2r-1}} \binom{2r-1}{r-1}$$

To estimate the transition time between the semi-stable states ({0, $R_2$} and {$R_2$, $R_2$}), we approximate the 2 dimensional dynamics as a 1 dimensional flow along the horizontal line, $R_2$ = 0.65, see Figure 4. We assume that $R^2$ varies only slightly around its stable state and that all variations in $R^2$ are due to noise processes affecting the dynamics of the cortical flow.

$$R^2 = R_s + r_2$$

We then obtain the following for the dynamics of $R^1$.

$$\frac{d}{dt}R^1_{i,0,1} = \omega_i g_{ii} \sum_{r=1}^{\infty} B_r \frac{{R^1_{i,0,0}}^{2r-1}}{2^{2r-1}} \binom{2r-1}{r-1} + \omega_i h_{ii} \cos(\varphi^1_{i,0,0}$$
$$- \varphi^2_{i,0,0})r_2 \sum_{r=1}^{\infty} B_r (2r-1) \frac{R_s^{2r-2}}{2^{2r-1}} \binom{2r-1}{r-1} + \omega_i h_{ii} \cos(\varphi^1_{i,0,0}$$
$$- \varphi^2_{i,0,0})r_2{}^2 \sum_{r=2}^{\infty} B_r \binom{2r-1}{2} \frac{R_s^{2r-3}}{2^{2r-1}} \binom{2r-1}{r-1}$$

To estimate the transition rate, we replace the actual displacement of the trajectory — in the $R^2$ direction — by the average variation over all possible trajectories, i.e.

$$r_2 \rightarrow \langle r_2 \rangle$$

$$r_2{}^2 \rightarrow \langle r_2{}^2 \rangle$$

Note that the averages of the powers of $r_2$ over all pathways depends on the noise process acting on the cortical column, or equivalently the (local) temperature of the system.
We then get the following,

$$\frac{d}{dt}R^1_{i,0,1} = \omega_i g_{ii} \sum_{r=1}^{\infty} B_r \frac{{R^1_{i,0,0}}^{2r-1}}{2^{2r-1}} \binom{2r-1}{r-1} + \omega_i h_{ii} \cos(\varphi^1_{i,0,0}$$

$$- \varphi^2_{i,0,0}) \langle r_2^2 \rangle \sum_{r=1}^{\infty} B_r \binom{2r-1}{2} \frac{{R_s}^{2r-3}}{2^{2r-1}} \binom{2r-1}{r-1}$$

$$A_1 \equiv \langle r_2^2 \rangle \sum_{r=1}^{\infty} B_r \binom{2r-1}{2} \frac{{R_s}^{2r-3}}{2^{2r-1}} \binom{2r-1}{r-1}$$

$$\frac{d}{dt}R^1_{i,0,1} = \omega_i g_{ii} \sum_{r=1}^{\infty} B_r \frac{{R^1_{i,0,0}}^{2r-1}}{2^{2r-1}} \binom{2r-1}{r-1} + \omega_i h_{ii} \cos(\varphi^1_{i,0,0} - \varphi^2_{i,0,0}) A_1 \qquad \text{Eq. S1.1}$$

This is a gradient flow where the corresponding potential is given by,

$$U = -\omega_i g_{ii} \sum_{r=1}^{\infty} B_r \frac{{R^1_{i,0,0}}^{2r}}{r 2^{2r}} \binom{2r-1}{r-1} - \omega_i h_{ii} \cos(\varphi^1_{i,0,0} - \varphi^2_{i,0,0}) A_1 R^1_{i,0,0} \qquad \text{Eq. S1.2}$$

## S2. A parametrization of the flow between meta-stable states.

The flow between the two states will be given by the following Fokker Planck equation,

$$\frac{\partial \rho}{\partial t} = -\frac{\partial}{\partial x}\left(\rho \left(\frac{\partial}{\partial x} U\right)\right) + \frac{\sigma_n^2}{2} \frac{\partial^2 \rho}{\partial x^2}$$

This flow is projected onto the space of normalized distributions, and parameterized by the mean (μ) and standard deviation (σ). To simplify the integration, we assume integration over the whole real line.

$$\int x(\rho + \partial \rho) - \int x\rho = d\mu$$

$$\frac{d\mu}{dt} = \int x \frac{\partial \rho}{\partial t}$$

$$\int x^2 \, \partial \rho - 2 \int x \partial \rho = (\sigma + d\sigma)^2 - \sigma^2 = 2\sigma d\sigma$$

$$\int (x^2 - 2x) \frac{\partial \rho}{\partial t} = \frac{2\sigma d\sigma}{dt}$$

The last term of the Fokker Planck eq. — the diffusion component — will not affect the mean but will have a component for the change in standard deviation.

$$\int (x^2 - 2x) \frac{\partial \rho}{\partial t} = \int x^2 \frac{\partial \rho}{\partial t} = \frac{2\sigma d\sigma}{dt}$$

$$\frac{\sigma_n^2}{2} = \frac{\sigma d\sigma}{dt}$$

$$\sigma^2 = \sigma_n^2 t$$

The drift term of the FP-eq gives the following,

$$-\frac{\partial}{\partial x}\left(\rho \frac{\partial}{\partial x} U\right) = \omega_i g_{ii} \sum_{r=1}^{\infty} B_r \left[\rho(2r-1)\frac{x^{2r-2}}{2^{2r-1}} + \frac{d\rho}{dx}\frac{x^{2r-1}}{2^{2r-1}}\right]\binom{2r-1}{r-1} + \omega_i h_{ii} \frac{d\rho}{dx} \cos(\varphi^1_{i,0,0} - \varphi^2_{i,0,0}) A_1$$

$$\frac{d\mu}{dt} = \int x \frac{\partial \rho}{\partial t} = \int \omega_i g_{ii} \sum_{r=1}^{\infty} B_r \left[\rho(2r-1)\frac{x^{2r-1}}{2^{2r-1}} + \frac{d\rho}{dx}\frac{x^{2r}}{2^{2r-1}}\right]\binom{2r-1}{r-1}$$
$$+ \omega_i h_{ii} x \frac{d\rho}{dx} \cos(\varphi^1_{i,0,0} - \varphi^2_{i,0,0}) A_1$$

$$\rho = \frac{1}{\sigma\sqrt{2\pi}} \exp\left(-\frac{1}{2}\left(\frac{x-\mu}{\sigma}\right)^2\right)$$

$$\frac{d\rho}{dx} = -\frac{1}{\sigma^2\sqrt{2\pi}}\left(\frac{x-\mu}{\sigma}\right)\exp\left(-\frac{1}{2}\left(\frac{x-\mu}{\sigma}\right)^2\right)$$

We now change the variable of integration as follows,

$$z = \frac{x-\mu}{\sigma}$$

$$\sigma dz = dx$$

$$dx\rho(x) = \frac{dz}{\sqrt{2\pi}} \exp\left(-\frac{1}{2}z^2\right) = dzp(z)$$

$$dx\frac{d\rho}{dx} = -\frac{dz}{\sigma\sqrt{2\pi}} z\exp\left(-\frac{1}{2}z^2\right)$$

$$\frac{d\mu}{dt} = \int \omega_i g_{ii} \sum_{r=1}^{\infty} B_r \left[ \rho(2r-1) \frac{(\sigma z + \mu)^{2r-1}}{2^{2r-1}} + \frac{d\rho}{dx} \frac{(\sigma z + \mu)^{2r}}{2^{2r-1}} \right] \binom{2r-1}{r-1}$$
$$+ \omega_i h_{ii} (\sigma z + \mu) \frac{d\rho}{dx} \cos(\varphi^1_{i,0,0} - \varphi^2_{i,0,0}) A_1$$

$$\int dz \rho (\sigma z + \mu)^{2r-1} = \sum_{s=0}^{r-1} \binom{2r-1}{2s} \sigma^{2s} \mu^{2r-1-2s} \int z^{2s} \rho$$

$$\left( \int z^{2s} \rho = (2s-1)!! \right)$$

$$\int dz \rho (\sigma z + \mu)^{2r-1} = \sum_{s=0}^{r-1} \binom{2r-1}{2s} \sigma^{2s} \mu^{2r-1-2s} (2s-1)!!$$

$$\int dx \frac{d\rho}{dx} (\sigma z + \mu)^{2r} = -\frac{1}{\sigma} \int dz z \rho (\sigma z + \mu)^{2r} = -\sum_{s=1}^{r} \frac{1}{\sigma} \binom{2r}{2s-1} \sigma^{2s-1} \mu^{2r-2s+1} \int z^{2s} \rho$$

$$\int dx \frac{d\rho}{dx} (\sigma z + \mu)^{2r} = -\sum_{s=1}^{r} \binom{2r}{2s-1} \sigma^{2s-2} \mu^{2r-2s+1} (2s-1)!!$$

$$\int dx (\sigma z + \mu) \frac{d\rho}{dx} = -\frac{1}{\sigma} \int dz\, z \rho (\sigma z + \mu) = -\int dz\, z^2 \rho = -1$$

We then have for the full integral,

$$\frac{d\mu}{dt} = \omega_i g_{ii} \sum_{r=1}^{\infty} B_r \left[ (2r-1) \frac{\sum_{s=0}^{r-1} \binom{2r-1}{2s} \sigma^{2s} \mu^{2r-1-2s} (2s-1)!!}{2^{2r-1}} \right.$$
$$\left. - \frac{\sum_{s=1}^{r} \binom{2r}{2s-1} \sigma^{2s-2} \mu^{2r-2s+1} (2s-1)!!}{2^{2r-1}} \right] \binom{2r-1}{r-1}$$
$$- \omega_i h_{ii} \cos(\varphi^1_{i,0,0} - \varphi^2_{i,0,0}) A_1$$

To simplify: if $h_{ii} > g_{ii}$ we get the following,

$$\frac{d\mu}{dt} = -\omega_i h_{ii} \cos(\varphi^1_{i,0,0} - \varphi^2_{i,0,0}) A_1 \propto -h_{ii} \cos(\varphi^1_{i,0,0} - \varphi^2_{i,0,0}) \qquad \text{Eq. S2.1}$$